\documentclass[10pt, conference, compsocconf]{IEEEtran}

\IEEEoverridecommandlockouts
\usepackage{cite}
\usepackage{amsmath,amssymb,amsfonts}
\usepackage{graphicx}
\usepackage{textcomp}
\usepackage{xcolor}
\usepackage{algpseudocode}
\usepackage{algorithm}
\usepackage{enumitem}
\usepackage{flushend} 

\usepackage{url}  
\makeatletter 
\g@addto@macro{\UrlBreaks}{\UrlOrds}  
\makeatother

\usepackage{soul}
\usepackage{caption}
\makeatletter
\let\MYcaption\@makecaption
\makeatother

\usepackage[font=footnotesize]{subcaption}

\makeatletter
\let\@makecaption\MYcaption
\makeatother

\usepackage{multirow}
\usepackage[bookmarks=false]{hyperref}
\usepackage{pgf, tikz}
\usetikzlibrary{arrows, automata}
\allowdisplaybreaks  

\newcommand{\xkji}[3]{\ensuremath{x_{#1,#2}^{#3}}}
\newcommand{\skj}[2]{\ensuremath{s_{#1,#2}}}
\newcommand{\ykjr}[3]{\ensuremath{y_{#1,#2}^{#3}}}
\newcommand{\pkj}[2]{\ensuremath{P_{#1,#2}}}

\algdef{SE}[SUBALG]{Indent}{EndIndent}{}{\algorithmicend\ }%
\algtext*{Indent}
\algtext*{EndIndent}

\newtheorem{problem}{Problem}
\newtheorem{theorem}{Theorem}

\newcommand{\sep}[1]{}

\newcommand{\squeezeuppicture}{\vspace{-3mm}} 

\begin{document}

\title{Skedulix: Hybrid Cloud Scheduling for Cost-Efficient Execution of Serverless Applications
}

\author{
\IEEEauthorblockN{Anirban Das, Andrew Leaf, Carlos A. Varela, Stacy Patterson\\}
\IEEEauthorblockA{
\textit{Department of Computer Science} \\
\textit{Rensselaer Polytechnic Institute, Troy, NY USA} \\
dasa2@rpi.edu, leafa@rpi.edu, cvarela@cs.rpi.edu, sep@cs.rpi.edu}
}

\maketitle

\begin{abstract}
We present a framework for scheduling multi-function serverless applications over a hybrid public-private cloud. 
A set of serverless jobs is input as a batch, and the objective is to schedule function executions over the hybrid platform to minimize the cost of public cloud use, while completing all jobs by a specified deadline.  As this scheduling problem is NP-Hard, we propose a greedy algorithm that dynamically determines both the order and placement of each function execution using predictive models of function execution time and network latencies. 
We  present a prototype implementation of our framework that uses AWS Lambda  and OpenFaaS,
for the public and private cloud, respectively. 
We evaluate our prototype in live experiments using a mixture of compute and I/O heavy serverless applications. 
Our results show that our framework can achieve a speedup in batch processing of up to 1.92 times that of an approach that uses only the private cloud, at 40.5\% the cost of an approach that uses only the public cloud.


\end{abstract}

%

\section{INTRODUCTION}
Serverless computing is an emerging paradigm for the deployment of application and services~\cite{jonas2017occupy, servermix, baldini2017serverless}. A serverless application consists of a collection of stateless functions that transfer information between them using platform services, such as databases and object stores. 
The serverless paradigm provides benefits including ease of development and deployment, as well as seamless elasticity, since an application can be scaled up by  executing multiple copies of the same function in parallel.

In public cloud serverless platforms, such as AWS Lambda~\cite{awslambda}, 
Azure Functions~\cite{azurefunctions}, the cloud provider itself maintains the platform and its underlying infrastructure. Each function is executed in its own sandboxed container, and the platform can scale up the application to meet demand by simply deploying more containers.  The application owner pays for the time during which the functions execute, as well as for any services the functions use.

Serverless applications can also be  deployed in a private cloud, i.e., a set of servers that is maintained by the application owner, using frameworks like OpenFaaS~\cite{openfaas} and Apache Openwhisk~\cite{openwhisk}. This private cloud may consist of privately owned hardware that is kept on the application owner's premises or on leased hardware in offsite premises. 
A private cloud offers several benefits to application owners. Sensitive data can be processed on protected servers without sending them to the public cloud, thus preserving user privacy~\cite{wang2018deep,varghese2018next}. Further, by processing jobs on site, the end-to-end application latency may be reduced. Finally, a private cloud offers cost benefits, especially for long-running jobs and services. However, because a private cloud has limited resources, 
it may not be possible to meet service level agreement (SLA) requirements for all workloads using the private cloud alone. 

To leverage the benefits from both the private and public clouds, an application owner can use a \emph{hybrid cloud} approach~\cite{varghese2018next}. 
In this model, an application owner maintains a private cloud that is resourced, for example, to process typical workloads with the desired performance guarantees. When the workload increases beyond the capacity of the private cloud, to maintain quality of service, some jobs, or parts of jobs, can be dispatched to the public cloud, with an incurred cost. 
This approach gives an application owner the privacy, performance, and cost benefits of a private cloud, while eliminating the need to maintain hardware for peak workloads that would otherwise go unused. The challenge is then to determine which jobs to run in the private cloud and which to run in the public cloud to achieve both high performance and low cost.

%
%
%
We propose Skedulix, a framework for scheduling multi-function serverless applications over a hybrid cloud. 
We consider a batch input scenario, where the workload consists of a set of serverless jobs, and the objective is to schedule function executions over the hybrid platform to minimize the cost of public cloud use, while meeting a user-specified deadline. 
A formalization of this scheduling and assignment problem can realized as a Mixed Integer Linear Program (MILP) and can be proved to be 
$\mathcal{NP}$-hard. Therefore, we propose a greedy approach, with two variations for cost minimization. Further, we implement and evaluate our approach in live experiments.

Specifically, our contributions are as follows. (i) We formalize a framework for scheduling multi-function applications in the hybrid cloud setting to minimize the cost of execution in public cloud. (ii) We then propose 
a greedy algorithm that dynamically 
determines both the order and placement of each function in the hybrid cloud. 
(iii) For scheduling, our algorithm requires accurate models of function execution time, data transfer times, and intermediate data sizes. We demonstrate how to generate such application-specific models from training data obtained using AWS Lambda in the public cloud and OpenFaaS in our private cloud. (iv) We present a prototype implementation of our framework that runs within this hybrid cloud platform. (v) We then evaluate the performance of our algorithm on this prototype using three canonical examples with mixed compute and I/O heavy workloads. 
Our results show that our framework can achieve a speedup in batch processing of up to 1.92 times that of an approach that uses only the private cloud, at 40.5\% the cost of an approach that uses only the public cloud.


The remainder of this paper is organized as follows. Sec.~\ref{model.sec} describes the platform model, outlines the problem we address, and describes the necessary elements for scheduling.
In Sec.~\ref{sec.scheduler_design_and_approximation_algorithm}, we present the scheduler design and algorithm.
In Sec.~\ref{sec.system_implementation}, we give the details of our framework implementation, including how we generate our performance models. Sec.~\ref{experiments.sec} presents an experimental evaluation of our framework. Sec.~\ref{related.sec} summarizes related work, and we conclude in Sec.~\ref{conclusion.sec}. 


\section{System Overview} \label{model.sec}
\begin{figure}[!t]
    \centering
    \includegraphics[scale=.45]{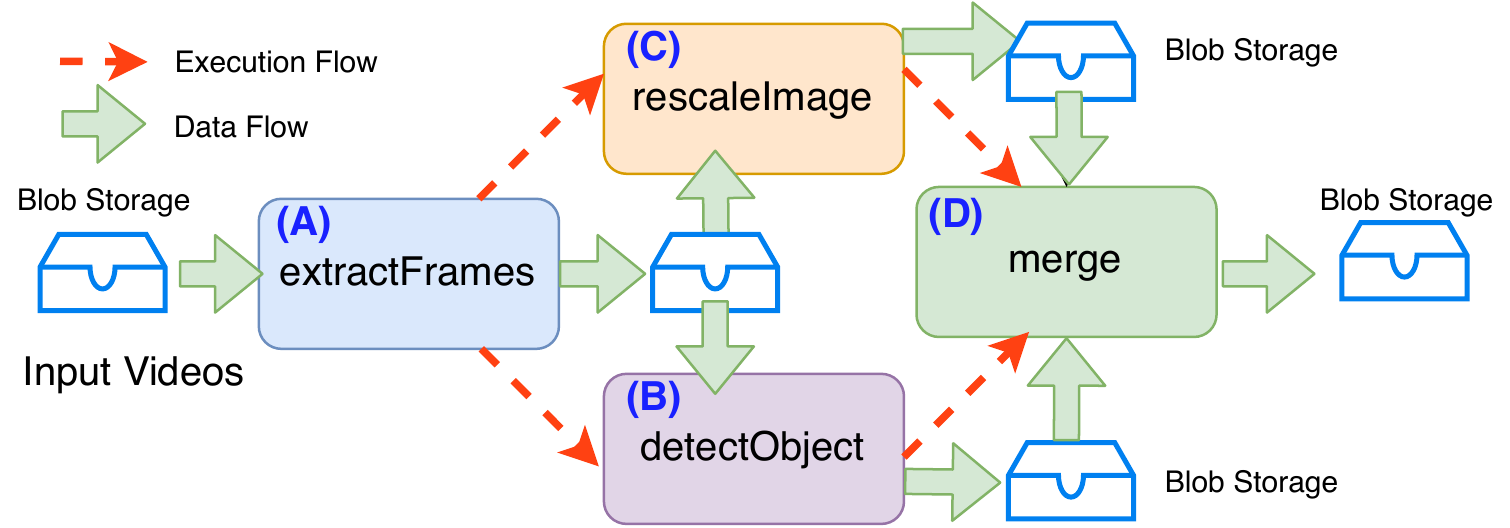}
    \caption{An example DAG representing precedence constraints in a Video Processing application.}
    \label{fig:NLfunctionchain_public}
    \squeezeuppicture
\end{figure}

\subsection{System Model}
We model applications as directed acyclic graphs (DAGs); each node is a function, or a \emph{stage}, and the DAG identifies the partial order in which the stages must execute. 
A sample DAG for a serverless Video Processing application is shown in Fig.~\ref{fig:NLfunctionchain_public}, where the red arrows represent precedence constraints (but no conditionals) of the stages in the workflow. For example, stage $B$ starts after stage $A$, and stage $D$ starts after both stages $C$ and $B$ complete.

We consider a hybrid cloud setting, where there are serverless platforms in both the private and the public cloud.
In the private cloud, we assume that there is a fixed number of \emph{replicas} allocated to each stage. For a stage $k$, we denote the number of deployed replicas by $I_k$. 
Each replica can execute a single function at a time, so that $I_k$ functions of stage $k$ may execute in parallel in separate containers, where the containerization provides resource isolation. 
The private cloud contains a storage service to store inputs and results, as shown in Fig.~\ref{fig:NLfunctionchain_public}. 
We assume that the public cloud has unlimited capacity and thus executes functions in parallel as needed with no performance degradation. 
The public cloud also provides a storage service, and 
data may be transferred between private and public cloud storage if necessary.


We assume cost of executing a function in the private cloud to be zero. The public execution cost of a function (in USD) is determined from the compute latency of the public cloud function, using the AWS Lambda cost model :
\begin{align}
h(t) = 100 \times ceil\left(\frac{t}{100}\right) \times \frac{\mathcal{M}}{1024} \times \frac{0.00001667}{1000} \label{eqn.cost}
\end{align}
where $t$ is the execution time of the function in milliseconds, and $\mathcal{M}$ is the memory configuration in MB of the Lambda function. 
We note that our framework can be trivially extended to any cost model that is a deterministic function of the execution time.

\subsection{Problem Formulation} 
We consider a scenario where there is a batch workload of $J$ jobs to be processed. Each job corresponds to a single execution of a serverless application, i.e., a DAG of stages. The input data for the workload is available in the private cloud at the beginning. 
A stage $k$ of a job $j$ can execute in either the private or public cloud. 
The objective is to process all $J$ jobs with minimal cost and store the results in the private cloud by a given deadline $C_{max}$. This deadline is also referred to as the \emph{makespan} of the batch. 
Intuitively, private cloud resources should be used as much as possible to reduce cost. If there are not sufficient computing resources to meet the deadline, function executions can be offloaded to the public cloud. The challenge is to identify which function executions to offload so as to minimize cost while meeting the deadline.

To achieve this, we need a \textit{schedule}. We define a schedule to be a combination of both an \textit{order} of job executions at each stage as well as an \textit{assignment} of executions for each stage to either a replica or the public cloud.

This scheduling and assignment problem can be formalized as a MILP and can be fed into a standard solver such as Gurobi \cite{gurobi}.
However, this problem is $\mathcal{NP}$-hard, and hence it is not tractable to find an optimal solution for even moderately-sized workloads. Hence we develop a greedy scheduling algorithm.
A detailed formulation of the MILP and the proof of $\mathcal{NP}$-hardness is provided in the Appendix. 

In the next section, we describe our scheduler design and scheduling algorithm. The scheduler requires accurate estimates of the application execution latencies in both the private and public clouds, as well as the data transfer time between them. We give the details of how we create models to generate these estimates in Sec.~\ref{subsec:perfmodel}. 

\section{Scheduler Design and Scheduling Algorithms} \label{sec.scheduler_design_and_approximation_algorithm}

\subsection{Scheduler Overview} \label{sec.scheduler_design}
The scheduler is designed as a long running service inside the private cloud. On receipt of a batch job execution request, the scheduler uses the algorithm described in the next section to decide where to run the stages of each job and in which order. The scheduler has one process for each stage in the application, as shown in Fig.~\ref{fig:system_architecture}. 
Each process maintains a queue of uncompleted jobs for its corresponding stage. 
When a private cloud replica for that stage is available, the scheduler process dequeues the job at the head of the queue at dispatches it to that replica. 
We explore several priority orders for the queue, designed to minimize the cost of public cloud utilization. We detail these orders in Sec.~\ref{subsec.heuristics}.

The scheduler process also monitors the  queue and offloads jobs to the public cloud when necessary to meet the makespan deadline.
When a private cloud replica completes executing a job stage, the job is added to the queue(s) of the next stage(s) in the application, as specified by the application's DAG.
If a job stage is offloaded to the public cloud, the downstream stages for the job are also executed in the public cloud.


\subsection{Scheduling Algorithm} \label{sec.approximation_algorithm}

The pseudocode for our scheduling algorithm is described in Alg~\ref{alg.main_algorithm}. 
We consider an application with $M$ stages, Let $\Phi$ denote the set of all stages in the application, and let ${\cal J}$ denote the input set of $J$ jobs.
 Let the batch start executing at time $t_0$. We let $P_{k,j}^{private}$ denote the estimated latency of executing stage $k$ for job $j$ in the private cloud and $P_{k,j}^{public}$ denote the estimated latency of executing the same stage for job $j$ in the public cloud. 

\begin{algorithm}[H]
\caption{Scheduling algorithm} \label{alg.main_algorithm}
\begin{algorithmic}[1]
\State Initialization:
\Indent
\State $\textstyle T_{max} = \sum_{k \in \Phi} I_k \times C_{max} $    \label{line.sort_start} 
\For{$j \in {\cal J}$}
		\State $\mathcal{C}_j = \textstyle \sum_{k \in \Phi} \pkj{k}{j}^{private}$ 
	\EndFor
	\State $\mathcal{C}_{all} = \{C_j \; |\; j=1,2\cdots, J\}$ sorted in priority order \label{line.sort_initial_list}

	\State $\textstyle \Upsilon = \textstyle \{j~|\; \sum_{j=1}^{q}\mathcal{C}_j \leq T_{max}\}$ 
	\State Dispatch jobs in ${\cal J} \setminus \Upsilon$ to public cloud 
	\State ${\cal J} = {\cal J} \setminus \Upsilon$ 
	\State Put jobs in ${\cal J}$ in priority queue(s) for first stage(s) \label{line.sort_end}
\EndIndent
\State At each stage $\ell$:
		\Indent
		\State \textbf{on} replica availability:
		\Indent
			\State Dequeue head of $\mathcal{Q}_{\ell}$ and dispatch to replica
		\EndIndent 
			\State \textbf{on} add or remove from $\mathcal{Q}_{\ell}$: \label{line.acd_start} 
		\Indent
			\State Make a copy $Q_c$ of $Q_\ell$ to loop over
			\For {job $j \in 1, 2, ... |Q_c|$ }			
				\State If $ACD_{\ell, j} <0$ 
				\State ~~~~~ dequeue $j$ and dispatch to public cloud
			\EndFor 
			\State Sync $Q_l$ with $Q_c$ \label{line.acd_stop}
		\EndIndent
		\EndIndent

\end{algorithmic}
\end{algorithm}

In initialization phase, before executing any job, we first get a rough estimate of the number of jobs that can fit in the computing capacity of the private cloud, and we offload any jobs that cannot fit within this computing capacity to the public cloud.
We calculate this computing capacity, $T_{max}$, as
\[
T_{max} = \sum_{k \in \Phi} I_k \times C_{max}, 
\]
i.e., $T_{max}$ is the total computing time in the private cloud if all replicas in all stages $k \in \Phi$ execute for the entire makespan duration $C_{max}$.
We then find the total estimated runtime of each job $j$ in the private cloud as $\mathcal{C}_j =\sum_{k \in \Phi} P_{k,j}^{private}$. The scheduler immediately offloads a set of jobs $\Omega$, in priority order, until the sum of the execution times of the remaining jobs is less than or equal to $T_{max}$. The jobs are offloaded from the tail of the priority queue during initialization phase.
This initialization phase is shown in Lines~\ref{line.sort_start} - \ref{line.sort_end}.
We note that this initial offloading to the public cloud will not be sufficient to meet the deadline, since to process the jobs in ${\cal J} - \Omega$ in the private cloud would require 100\% utilization of all replicas for the entire duration until $t_0 + C_{max}$. This is not possible due to both the framework overhead and the stage precedence constraints. Therefore, we must offload additional job stages throughout the batch execution. We do this adaptively, as described next.

For each job stage $\ell$, its scheduling process monitors for jobs that it estimates may not complete by the deadline, and it offloads them to the public cloud. 
To determine whether a job $j$ will result in a violation, we use the \emph{apparent closeness to deadline} $ACD_{\ell,j}$.
The scheduler computes the estimated latency along the longest-latency path from the current stage to the final stage(s) and checks whether there is sufficient time remaining to execute all stages in this path for the job.
We denote the set stages along this path by $\Gamma(\ell)$.
At time $t$, the $ACD_{\ell,j}$ is computed as follows:
\[
ACD_{\ell,j}(t) = \mathcal{D} - (t + \sum_{y<j, y \in Q_\ell}\pkj{\ell}{y}^{private}/I_\ell +\sum_{k \in \Gamma(\ell)} \pkj
{k}{j}^{private})
\]
where $\mathcal{D}= t_0 + C_{max}$.
Here, $\mathcal{D}-t$ indicates the time remaining for execution before the deadline. The first summation is the estimated current queue delay. This is the sum of private cloud execution latencies of jobs prior to job $j$ in $Q_\ell$, divided equally among available replicas $I_\ell$.
This estimate holds under the assumption that the total work at stage $\ell$ is evenly distributed among all replicas. 
We also add $\sum_{k \in \Gamma(\ell)} \pkj{k}{j}^{private}$ to get an optimistic estimate about total time needed to complete job $j$ in the private cloud.
If $ACD_{\ell,j}$ is negative, we estimate that job $j$ will not complete by the deadline, and thus should be offloaded to the public cloud.


Whenever there is a change in the priority queue $Q_\ell$ for a stage, i.e., a job is added or removed, 
the value of $ACD_{\ell,j}$ is computed for all the jobs $j$ inside the queue by looping over a local copy ($Q_c$) of $Q_\ell$. If $ACD_{\ell,j} < 0$, the job is dispatched to the public cloud.
These steps are shown in Lines~\ref{line.acd_start} - \ref{line.acd_stop}. Finally, we update $Q_\ell$ with jobs in $Q_c$ in the correct order.
A job remains in the priority queue until it is dispatched to a replica in the private cloud or until its $ACD$ becomes negative and it is offloaded to the public cloud.
Since these steps are executed at every stage, this approach allows us to offload enough jobs to the public cloud, in priority order, such that the remaining jobs can finish in the private cloud by the deadline.

\subsection{Priority Queue Sort Orders} \label{subsec.heuristics}
The selection of which jobs to offload to the public cloud is made based on their order in the priority queue.
We consider the following priority orders.
\paragraph{\textbf{Highest Cost First order (HCF)}}
In this method, jobs are ordered
 by their cost of execution in the public cloud. 
 We can obtain the cost of execution of stage $k$ of job $j$ in public cloud from  $P_{k,j}^{public}$ using Eqn.~\ref{eqn.cost}. 
  Here, the less expensive jobs are offloaded to the public cloud first.

\paragraph{\textbf{Shortest Processing Time order (SPT)}} \label{subsec.description_SPT}
Here, we order jobs in Shortest Processing Time (SPT) order i.e. the jobs with lower processing time are always towards the head of the priority queue. We offload the jobs with longer processing time that are present towards the tail of the priority queue to the public cloud when necessary.
Since AWS rounds up the execution time of Lambda functions to the nearest 100 ms, we make an observation that, if we offload jobs with longer duration to the cloud, the rounding penalty will be a smaller fraction of the total cost. Hence, we waste less budget relatively for those jobs. 
Further, executing longer jobs in the public cloud can avail the benefit of parallelism without affecting the makespan of the entire batch of jobs negatively.

\section{Framework Implementation} \label{sec.system_implementation}
In this section, we describe the architecture of our framework, Skedulix, and we provide details of our prototype implementation. We also summarize how we generate the performance models used in our scheduling algorithm.

\subsection{Framework Architecture}

\begin{figure}[!t]
	\centering
	\includegraphics[scale=0.30]{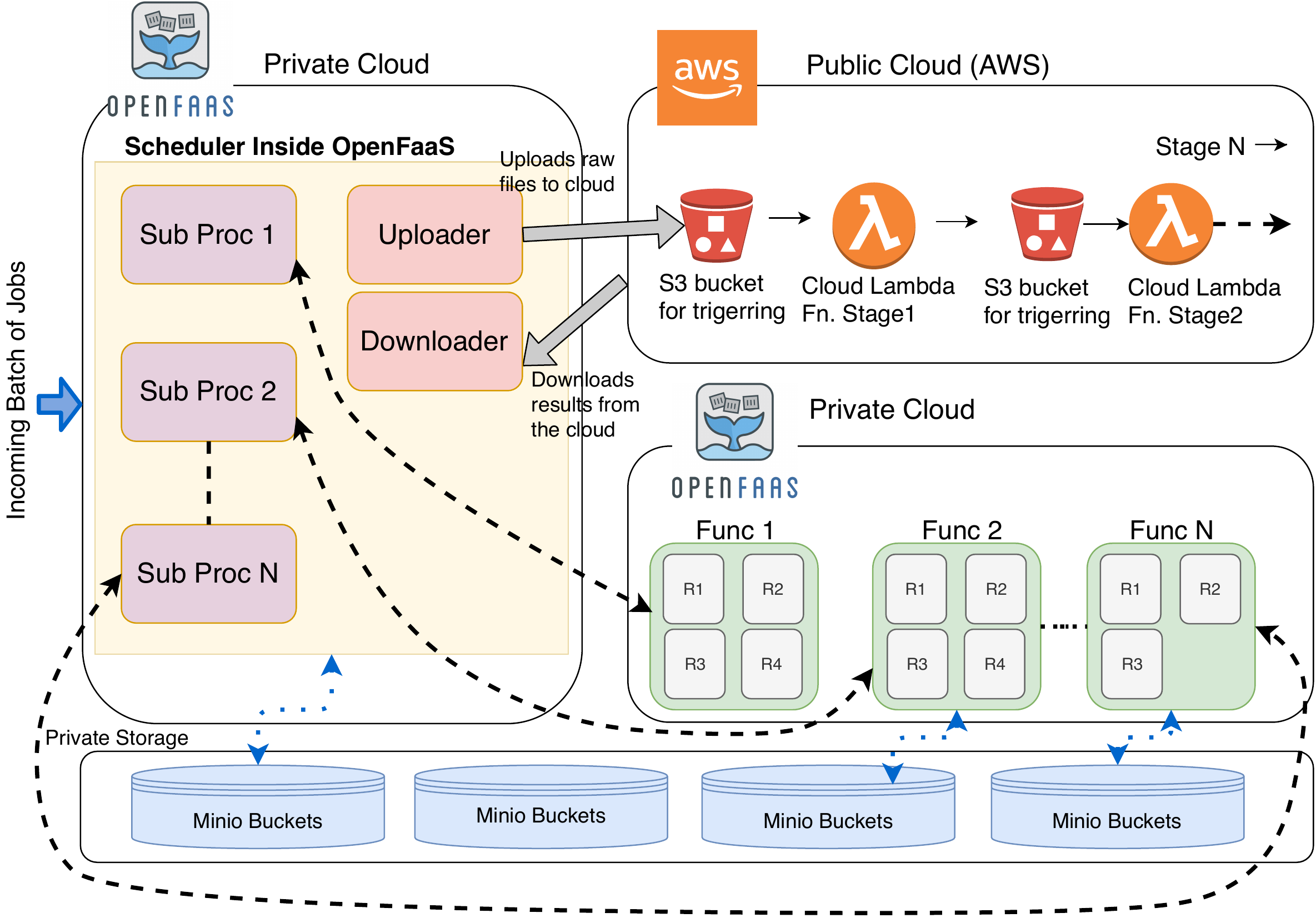}
	\caption{Skedulix framework architecture and the scheduler implementation on OpenFaaS.}
	\label{fig:system_architecture}
	\squeezeuppicture
\end{figure}

The framework architecture is shown in Fig.~\ref{fig:system_architecture}. For the public cloud deployment, we use AWS Lambda to execute functions corresponding to the stages, and we use AWS S3 to store inputs and intermediate results. 
To implement function chaining in AWS, we create unique S3 buckets that trigger their corresponding functions; one function triggers another by storing its output in the next function's input S3 bucket.
If at any stage, a job is scheduled to be executed in the public cloud, the corresponding scheduler process uploads the appropriate raw input to the specific S3 bucket in the public cloud. This raw input upload subsequently triggers the corresponding Lambda function for that stage. 
For DAGs with parallel stages, we use AWS Step Functions~\cite{awsstepfunctions}. 

For the private cloud platform, we use OpenFaaS as the serverless platform, and we use Minio~\cite{minio} for the storage. 
We deploy OpenFaaS on top of the Kubernetes Container Orchestration System \cite{kubernetes}. 
When a function is deployed in OpenFaaS, the latter initiates a function instance that runs inside a Kubernetes pod and also exposes an API address. The API address, when invoked, executes that deployed function. 
Since we need each replica of a function to be separately addressable by our custom scheduler, we configure our system such that each OpenFaaS function instance has exactly one pod at any time. 
We then create $I_k$ replicas for each function by creating $I_k$ versions of the same function, 
which can each be 
 uniquely addressed via simple \texttt{http} calls.

The input data for the batch workload is stored in a Minio bucket in the private cloud. 
The scheduler then uses Alg~\ref{alg.main_algorithm} to execute jobs in the private and public clouds. All interactions between function replicas and the scheduler are done via \texttt{http} requests. This includes the replica notifying scheduler of its availability and notifying the scheduler to execute downstream stages when a job stage completes. 
If the last stage of the application executes in the public cloud, the last function notifies the scheduler when it completes. The scheduler then downloads the results from the S3 bucket to a Minio bucket in the private cloud.

\subsection{Performance Modeling} \label{subsec:perfmodel}
To execute jobs in our framework using Alg.~\ref{alg.main_algorithm}, we need estimates for the application execution latencies in the private cloud ($P_{k,j}^{private}$) and the public cloud ($P_{k,j}^{public}$). 
To estimate these quantities, we train performance models using traces gathered from  executing a substantial number of jobs on AWS and in our private cloud. We use Python and \texttt{scikit-learn} library 
for training and subsequently tuning the models via cross-validation as described next.
We create separate models for each stage $k$ of an application where each model makes a prediction for each job $j$.

For the private cloud, for each stage $k$, we model the execution latency $P_{k,j}^{private}$ as the sum of the following components:  
\begin{enumerate}[leftmargin=*]
\item Function compute time: This is parameterized by function input properties, e.g., the file size or data dimension. 
\item Framework overhead: This is modeled using the mean over the training dataset. Its magnitude for each application stage is on the order of 15-20~ms. 
\end{enumerate}
For the public cloud, for each function, we model $P_{k,j}^{public}$ as 
a linear function of the input features (e.g., file size or data dimension). 

For all stages except the first stage of an application, we must also predict the size or properties of the input data for that stage, i.e., the size of the output data from the preceding function(s), since the latency performance models are parameterized by these properties.
For example, in the Video Processing application shown in Fig.~\ref{fig:NLfunctionchain_public}, to model the latency of the \texttt{detectObject} function, we need the characteristics of the output files of the \texttt{extractFrames} function. Therefore, for each function, we create a model that predicts the size of its output as a function of the size of its input.

For all of the above quantities, except the framework overhead, we  create our models using regularized ridge regression over the training data. 

\section{Experimental Evaluation} \label{experiments.sec}
In this section, we present an evaluation of our framework using three serverless applications. We first describe these applications, with the corresponding workloads and infrastructure setup in Sec.~\ref{sec.env_and_infra}. We describe the performance modeling and error in Sec.~\ref{model_accuracy.sec} and present the results from using our framework on these applications in Sec.~\ref{framework_experiments.sec}. 

\subsection{Experimental Environment and Infrastructure} \label{sec.env_and_infra}

\subsubsection{Serverless Applications} \label{apps.sec}
We employ three different applications that exhibit different canonical behaviors in terms of resource utilization. 
The first, the Matrix Processing application is compute-heavy, with minimal I/O. The second, Video Processing, consists of a mixture of compute-heavy and I/O bound functions. The third is the Image Processing application, which is I/O-heavy, with smaller compute requirements.

\textbf{Matrix Processing Application:} The application consists of two stages. The first stage is a matrix multiplication stage (MM). This stage takes, as input, a matrix in a CSV-formatted text file
and computes a product matrix by multiplying the matrix with its transpose. The resulting matrix is saved to storage in the same format.
The second stage (LU) takes in this product matrix, computes an LU decomposition, and  stores the results.
The input matrices are random integer matrices of dimensions between $350 \times 350$ and $500 \times 500$. This is a synthetic workload which we use as a characterization of a compute heavy extract-transform-load (ETL) workload. 

\textbf{Video Processing Application:} 
This is a Video Processing application with four stages, as shown in Fig.~\ref{fig:NLfunctionchain_public}. Key frames are first extracted from the input video clips using the \texttt{extractFrames} (EF) function, and the resulting images are stored in blob storage as zip. The \texttt{detectObject} (DO) function then loads the images from storage, detects the objects in each image and saves resulting coordinates and inference in a text file. The \texttt{rescaleImage} (RI) function rescales the images to lower resolution and saves the resulting images to storage in a single zip file. Finally, the \texttt{merger} (ME) function combines the results of the DO and RI functions into another zip file, and saves it to storage. This application is representative of a traffic surveillance application that detects objects in frames such as cars, buses, bikes etc. 

For our evaluation, we rescale each image to half its original resolution in the RI stage. 
Further, at the EF stage, we extract one key frame per second from the video files. For the input dataset, we use videos from the BDD100K database~\cite{yu2018bdd100k}. All input videos are of duration $<$10s in our experiments. 

\textbf{Image Processing Application:} This application has three stages. The input to this application is an image file of arbitrary size. The first stage is a \texttt{rotate} function, which rotates the image. The output is an image file of similar, but non-identical, size to the input.  The second stage is a \texttt{resize} function, which scales the image to a specific configurable size, which is $200\times200$ pixels for our experiments. While the number of pixels in the image is uniform across all output files of \texttt{resize} function, the output file size, in bytes, is not. Thus, our output size prediction models play a crucial role in the scheduling for this application. The final stage is a \texttt{compress} function, which reduces the quality of the image. 
For the input dataset, we use the open-source Image of Groups~\cite{gallagher_cvpr_09_groups}.


\subsubsection{Experimental Setup}
We set up OpenFaaS in a private cloud, consisting of a single node Kubernetes cluster on a machine with Intel Xeon CPU E5-1650 with 12 cores and 64 GB ram. The machine is connected to the internet via a wired LAN connection through 
a 1~Gbps network. It is synced with our private 
GPS time source for accurate time-keeping. 
The private object storage, Minio, is set up inside the same machine. 
Further, for the purpose of the experiments, in our private cloud, we use only two replicas to allow us to study the impact of limited resources. 
Accordingly, throughout all the experiments, we fix the CPU and RAM resources of replicas of all functions in the Matrix processing application at 1.0~CPU and 512~MB RAM, all function replicas in the Image processing application at 0.2~CPU and 512~MB RAM. For the Video processing application we configured each replica of EF at 0.5~CPU and 1024~MB RAM, DO at 1.0~CPU and 2048~MB RAM, RE and ME function replicas at  0.2~CPU and 512~MB RAM each.

For the public cloud, we use the US-East-1, North Virginia as the region of our AWS data center. For Matrix Multiplication and Image Processing, we set the memory of all the AWS Lambda functions at 2048 MB. For Video Processing, we set the memory of the EF and RI functions at 1024 MB, the DO function at 3008 MB, and the ME function at 512 MB. 
For this study, we consider warm starts only, and hence, we pre-warm a sufficient number of Lambda functions before doing the experiments.


We train the performance models as described in Sec.~\ref{subsec:perfmodel} for all applications. For the 
Matrix Processing application, we use 774 matrices for training and 150 as the test set for the live experiments. For Video Processing, we use 800 videos for training and 200 as the test set, and for Image Processing we use 800 images for training and 200 for the test set.  The regression models are selected  using the Grid Search method 
of \texttt{scikit-learn} on  the model parameters, with five-fold cross validation.

\subsection{Performance Model Accuracy} \label{model_accuracy.sec}
To study the accuracy of our performance models, we present results on the test set for each application.

\sep{It is not clear why you are reporting on all of these statistics, since we do not use them all in the scheduler. Let's discuss this section.}
\subsubsection{Matrix Processing Models}
We model the compute latency of the MM and LU stages as functions of the size of the input matrix, in bytes, and of the total number of entries in the input matrix, respectively. In the private cloud we obtain a Mean Absolute Percentage Error (MAPE) of 6.51\% and 4.57\%, respectively, for the MM and LU stages.  
The MAPE of the public cloud function execution latencies of MM and LU are 5.74\% and 2.52\%, respectively. 
The LU stage of the Matrix Processing application depends only on the dimensions of the input matrix, which can be determined from the input to MM stage. Thus, we do not need a prediction model for the intermediate data in this stage.

\subsubsection{Video Processing Models}
We model the compute latency of the Video Processing application as a function of the input filesize in bytes and the duration of the original video file. 
For the private cloud, we model the compute latency of the EF, DO, RI and ME stages and obtain a MAPE of 4.42\%, 1.44\%, 8.48\% and 51.3\% respectively. 
For the public cloud function execution latency, we observe a MAPE  of 5.28\%, 1.52\%, 7.69\% and 23.62\% for functions EF, DO, RI and ME, respectively.
The latency for the ME stage, which just merges the outputs of RI and DO, is very low in magnitude and does not show any clear pattern, resulting in high MAPE.
However due to the small magnitude, this error has low impact on the performance prediction for the entire application.
For the Video Processing application we need the output size prediction models from inputs for stages EF, RI and ME, MAPE of which are, respectively, 38.6\%, 5.24\% and 0.2\%.

\subsubsection{Image Processing Models} 
For the private cloud, in modeling compute latency of different stages, we observe a MAPE of 13.71\%, 12.24\%, and 12.91\%, respectively, for the Rotate, Resize, and Compress stages.
We further observe a MAPE of 26.1\%, 26.5\% and 29.5\% for the public cloud function execution latencies of the Rotate, Resize, and Compress functions. The latencies observed in the Image Processing application are of high variance and small magnitude, hence, we observe overall large error magnitudes. 
MAPE of prediction of output size from input of Rotate is 7.08\%, that of Resize is 11.69\% and that of Compress is 0.52\%.

\subsection{Scheduling Framework Evaluation} \label{framework_experiments.sec}
In this section, we present an experimental evaluation of our hybrid cloud scheduling framework prototype. 
In each experiment, the input batch arrives at the private cloud at time $t_0$. The makespan is counted from this starting time until the timestamp of the last saved file in the result bucket in Minio, after all jobs have completed. For each application, we explore a range of values of $C_{max}$. For the Matrix Processing application we explore $C_{max}$ between {300-700~s}; 
for Video Processing, we explore $C_{max}$ between 200-400~s, and for Image Processing, we use $C_{max}$ between 13-17~s. 

As points of comparison, we also include results for executions that take place entirely in the public cloud and entirely in the private cloud
For the all-public cloud execution, the batch of input files is uploaded in parallel to the input bucket of first stage in the public cloud. 
For the all-private cloud execution, we choose $C_{max}$ large enough and execute all jobs using SPT order.
 All experiment results are averaged over three runs, and error bars show one standard deviation.

For small scale experiments, we also compare the performance of our scheduling approach to the performance obtained from an optimal schedule.
To generate the optimal schedule, we need to model some additional latency constituents. This includes the public cloud function start-up latency and the data transfer latencies between the public and the private cloud. We model the start-up latency by taking the mean over the 99 percentile of measurements of the training data. We model upload and download latencies between the private and public cloud as a function of the data size, in bytes. We note that we generate application-specific models for these quantities based on the range of input and output sizes expected for each application using regularized ridge regression. 
We then solve the scheduling MILP using the Gurobi solver. We let the solver run for $>$20 hours, until we observe convergence of the objective function.

\begin{figure}[!t]
\centering
\begin{subfigure}[b]{.47\linewidth}
      \centering
      \includegraphics[width=\linewidth]{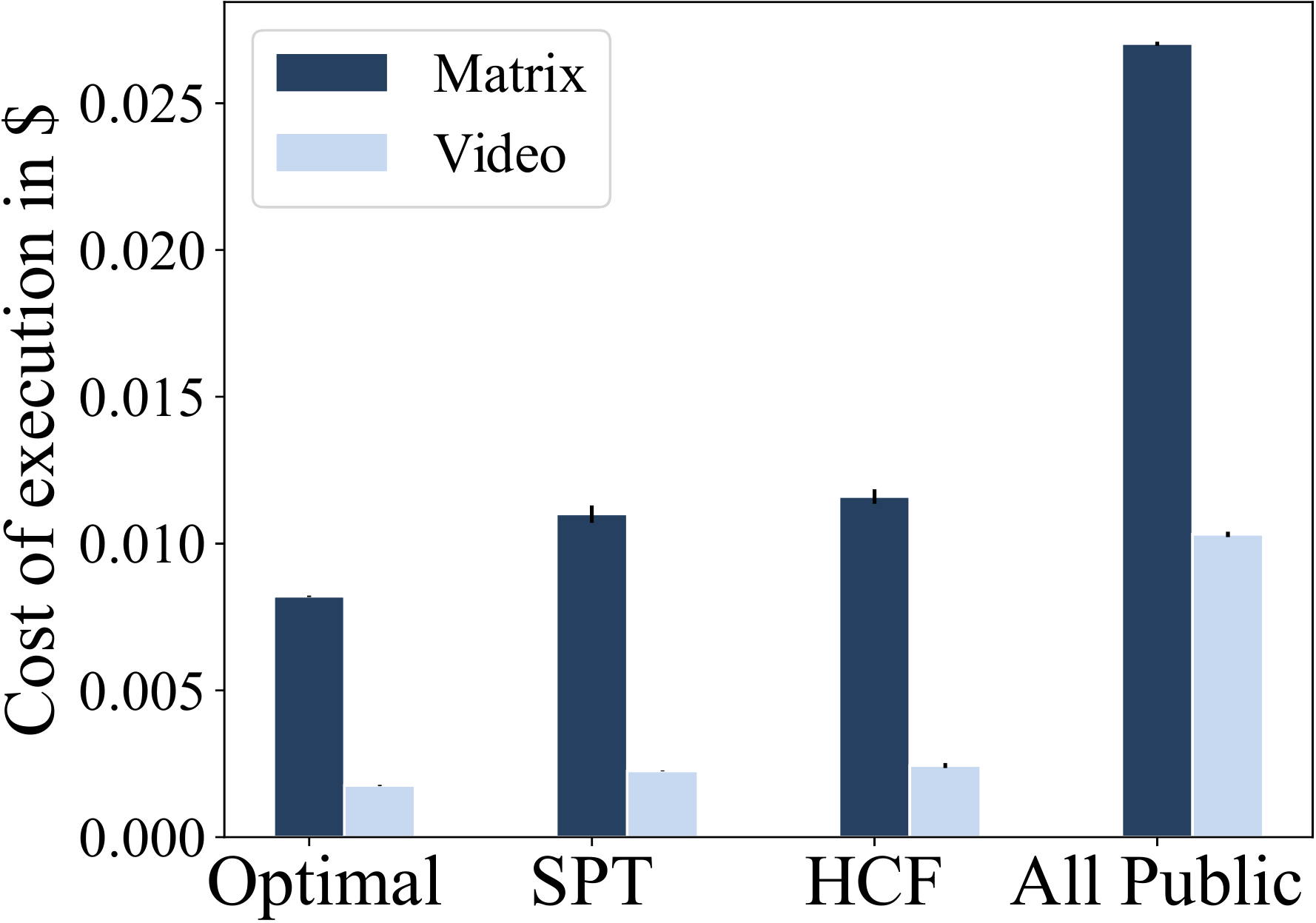}
      \caption{Total execution cost}      \label{fig.spt_gurobi_matrix_video_cost}
     \end{subfigure}
\begin{subfigure}[b]{.47\linewidth}
      \centering
      \includegraphics[width=\linewidth]{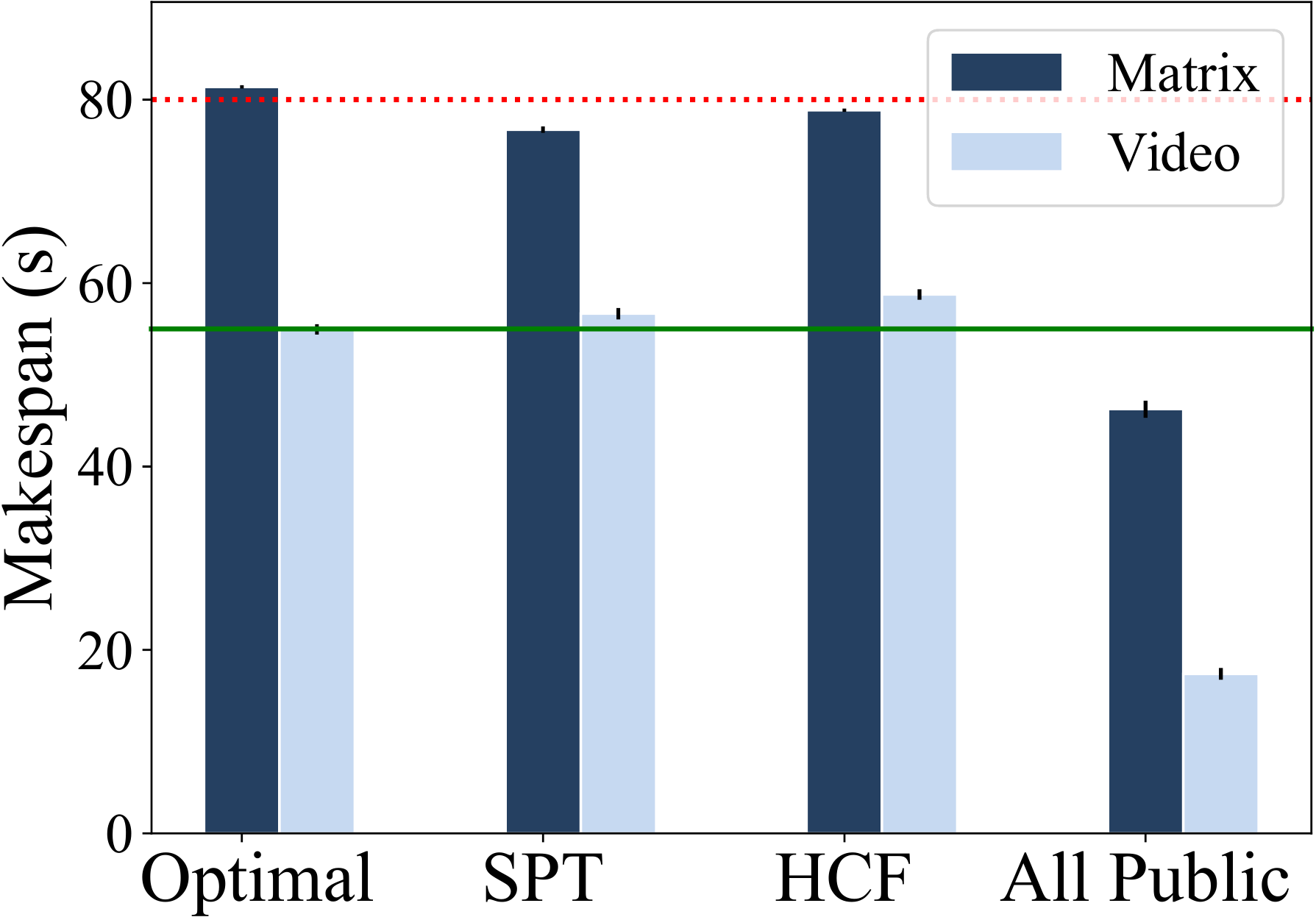}
      \caption{Makespan of execution}      \label{fig.spt_gurobi_matrix_video_makespan}
     \end{subfigure}
\caption{Comparison of the optimal, SPT and HCF schedules, and the all-public cloud execution, with 30 input jobs for the Matrix Processing and Video Processing Application.
}
\label{fig.spt_gurobi_matrix}
\squeezeuppicture
\end{figure}

\subsubsection{Optimal vs. SPT and HCF}


We first compare the performance of our scheduling algorithm, with the SPT and HCF priority orders, to the optimal schedule.
We study both the Matrix Processing and Video Processing applications. For each application, the input batch consists of 30 jobs, selected at random from the respective test sets.  
For the Matrix Processing application, we use $C_{max} = 80\;s$, and for the Video Processing application, we use $C_{max}=60\;s$.
We also show results for the all-public cloud execution.

Fig.~\ref{fig.spt_gurobi_matrix_video_cost} shows the total cost of execution for each application, and Fig.~\ref{fig.spt_gurobi_matrix_video_makespan} shows the actual makespan.
First, we observe that for the optimal, HCF, and SPT schedules, the makespan is very close to $C_{max}$, and in fact, the HCF and SPT schedules complete before $C_{max}$. The all-public cloud execution is much faster because of the parallelism offered by the cloud, but at the same time, it is much more expensive than HCF and SPT, showing the benefit of the hybrid cloud execution. The performance of the SPT and HCF heuristics are close to each other, however, HCF is slightly more expensive. Further, we observe that the SPT algorithm has 34\% higher cost than the run with the optimal schedule for the Matrix Processing application and 28.2\% higher cost than the optimal schedule for the Video Processing application. The greedy algorithm thus performs quite close to optimal, especially when we consider the fact that, obtaining the optimal schedule is infeasible for larger number of jobs.

\subsubsection{SPT vs. HCF with varying $C_{max}$} \label{subsubsec.SPT_HCF_meeting_deadlines}
We next study  the trend of execution cost and number of offloaded functions in SPT and HCF priority orders in our greedy algorithm with varying values of $C_{max}$. For the Matrix Processing application, we use all 150 jobs in the test set and for the Video Processing application we use all 200 jobs in the test set. We also use the Image Processing application to observe the performance of our heuristics on an application with smaller latencies. For this application, we use all 200 images in the test set.

\begin{figure*}[!t]
\centering
\begin{subfigure}[b]{.33\linewidth}
      \centering
      \includegraphics[width=.9\linewidth]{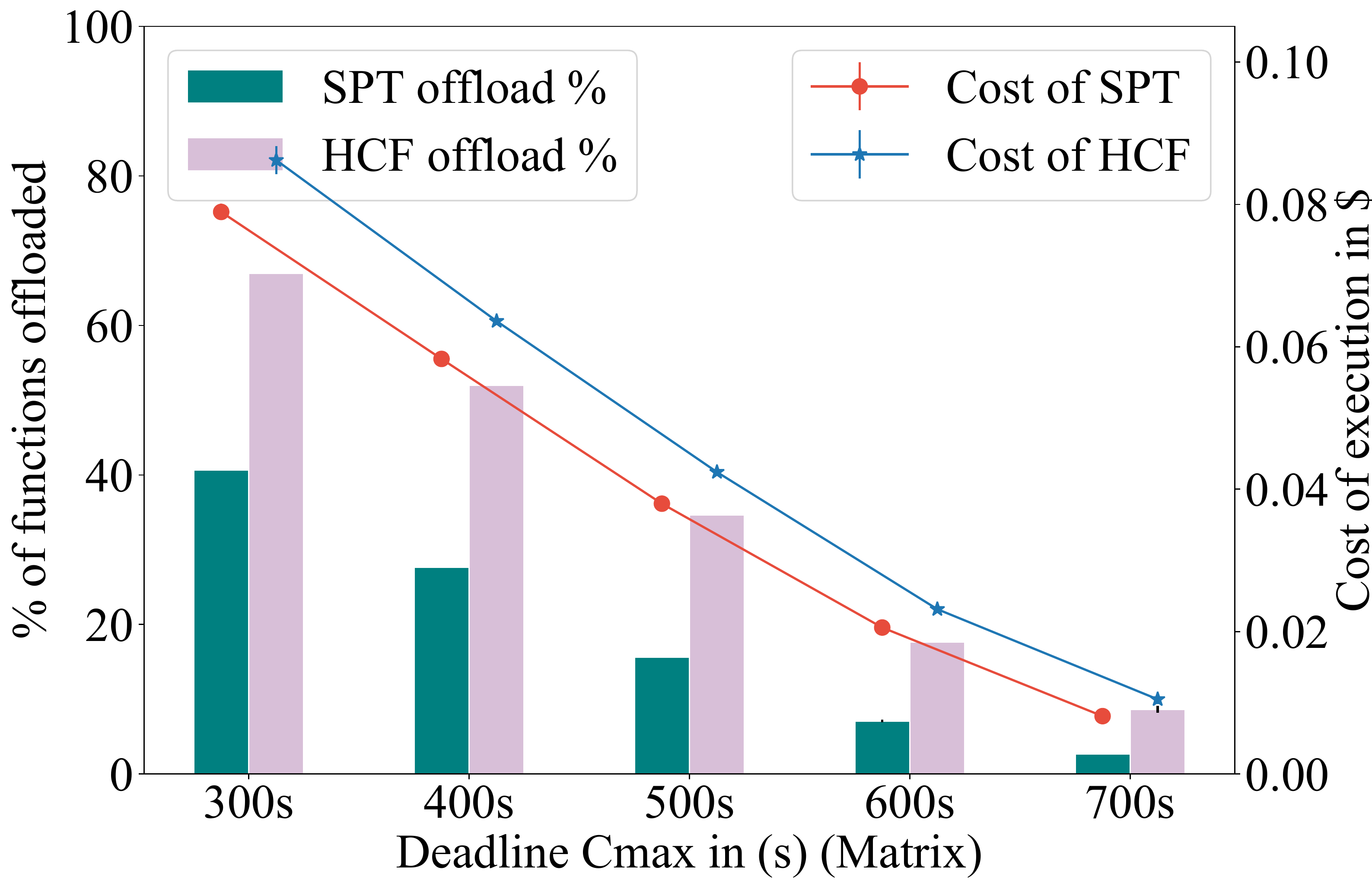}
      \caption{Matrix Processing (150 jobs, 300 function calls)}      \label{fig.all_jobs_matrix}
     \end{subfigure}%
\begin{subfigure}[b]{.33\linewidth}
      \centering
      \includegraphics[width=.9\linewidth]{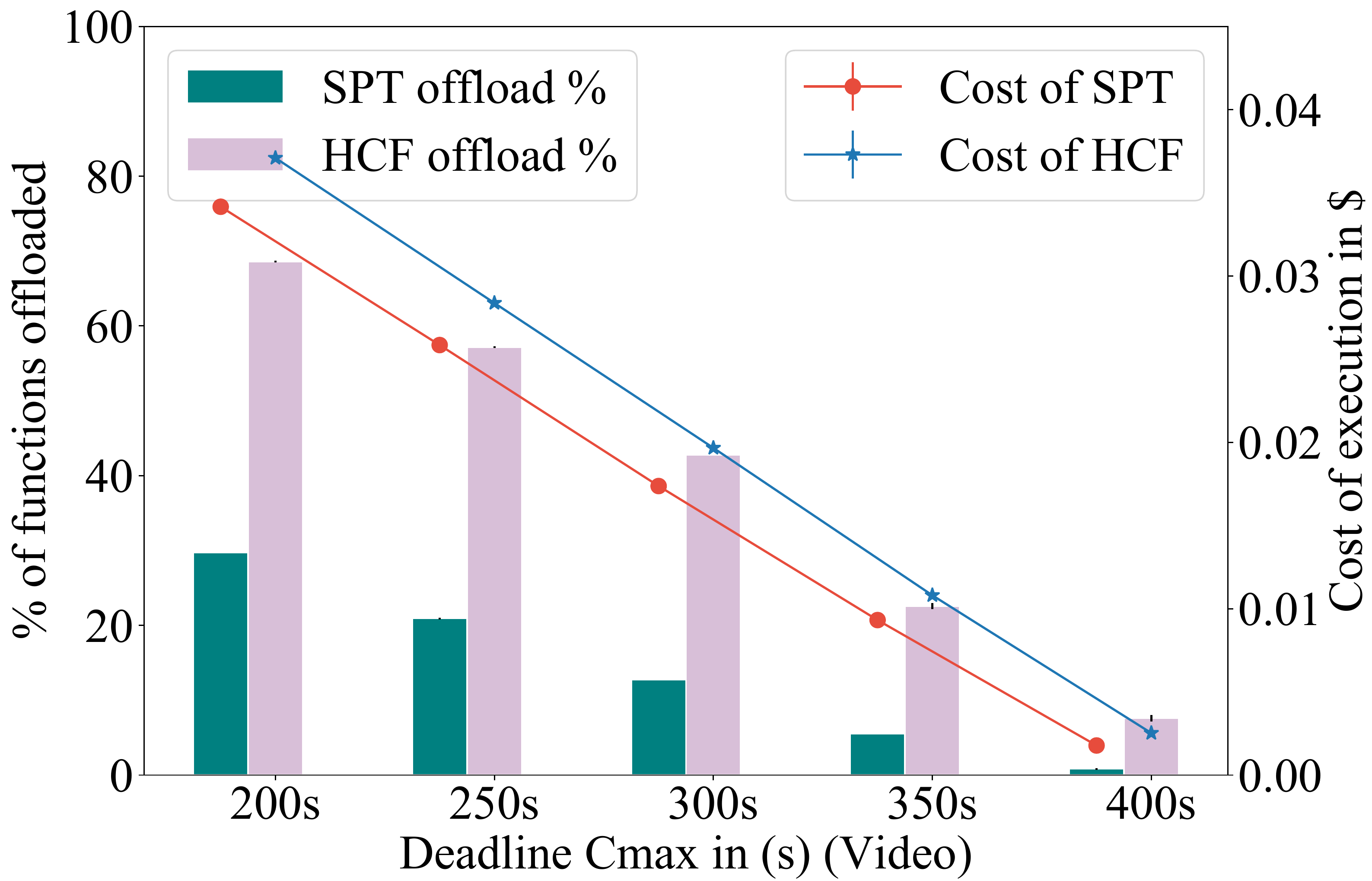}
      \caption{Video Processing (200 jobs, 800 function calls)}      \label{fig.all_jobs_video}
     \end{subfigure}%
\begin{subfigure}[b]{.33\linewidth}
      \centering
      \includegraphics[width=.9\linewidth]{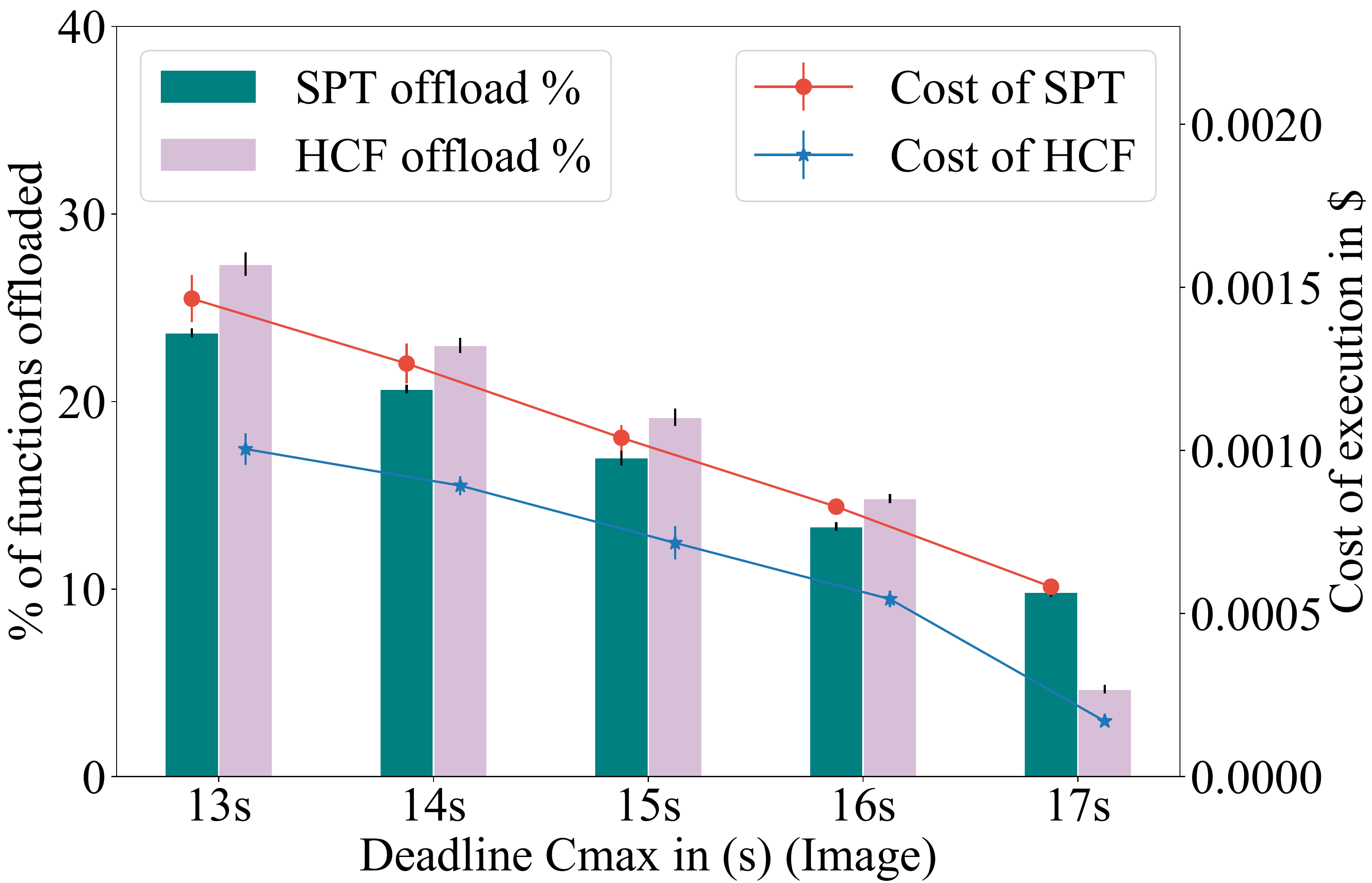}
      \caption{Image Processing (200 jobs, 600 function calls)}      \label{fig.all_jobs_image}
     \end{subfigure}
\caption{Comparison of the number of functions offloaded to the public cloud and the total execution cost for the SPT and HCF scheduling polices, with varying deadlines $C_{max}$, for the Matrix, Video and Image Processing applications. \sep{The fonts are too small in these figures. Can you make them larger? Also, can we come up with a shorter title for the left and right Y-axis?}}
\label{fig.all_jobs_cmax_variation}
\squeezeuppicture
\end{figure*}

The results are shown in Fig.~\ref{fig.all_jobs_cmax_variation}. The left Y-axis denotes percentage of the total number of function executions offloaded to the public cloud. The right Y-axis denotes the total public cloud execution cost.
We first observe that the total number of function stages offloaded to public cloud is a decreasing function of the deadline for both heuristics. Our scheduler offloads more job stages to the public cloud as the deadline decreases. This, in turn, increases the total cost of execution.

We observe that for all applications, the total number of function stages offloaded is higher for HCF compared to SPT priority order in general. This is because the HCF order tries to execute more expensive jobs in the private cloud, which roughly corresponds to many jobs with long durations in the private cloud. Hence, it ends up offloading a substantially larger number of inexpensive jobs of short durations.
\sep{I don't see why offloading longer jobs leads to offloading smaller jobs?}
This also results in higher overall cost, as seen in the Fig.~\ref{fig.all_jobs_cmax_variation}, for the Matrix and Video Processing applications.
\sep{I wouldn't bring up data costs, since we don't include these in our models. I don't think we should make this omission so obvious to the reviewers.}
On average, across all values of $C_{max}$, we see the trend that the HCF ordering is 14.3\% more expensive than SPT for the Matrix Processing application, and the HCF ordering is 17.9\% more expensive than SPT in the Video Processing application. However, the trend is the opposite in the case of the Image Processing application in Fig.~\ref{fig.all_jobs_image}. We see that the cost of the HCF heuristic is actually lower than SPT even when SPT offloads fewer jobs. Here, the number of functions offloaded in SPT and HCF are very close across different $C_{max}$. The number of offloaded jobs across SPT and HCF being very close, coupled with the fact that SPT offloads larger jobs, results in the total cost being higher for SPT than that of HCF in this case. 

In these applications, there can be bottleneck stage(s) where the private cloud execution latency of jobs are in general larger than the other stages. In order to maintain the makespan, and to keep the public cloud execution cost low, a good choice would be to prefer offloading the bottleneck stage(s) instead of the entire job. Our scheduler correctly offloads these stages to to the public cloud to meet the makespan. For the Matrix Processing application, this corresponds to the LU or the last stage. For Video Processing application this is the DO stage, and the scheduler correctly offloads the DO and ME function most frequently. Finally, for Image Processing, Rotate is the bottleneck stage, hence, all three functions gets offloaded to cloud once the scheduler decides to offload a job at Rotate. 

\sep{There are many observations in this paragraph. Can you give the reader any intuition here? I'm not sure what I am supposed to take away from this paragraph.} 

From these experiments, we observe that there is a clear trade off between performance, in terms of latency and cost.
Our hybrid cloud scheduling framework provides a mechanism for an application owner to determine their own balance of cost and performance by selecting the value for $C_{max}$.
Offloading longer jobs to the public cloud using SPT priority order works very well in practice in this system model for medium-high compute heavy workloads.

\sep{You need to explain what the experimental scenario is for this subsubsection. If it is the same as for the previous subsubsection, then perhaps we do not need a new subsubsection header?}

\begin{figure}[!t]
\centering
\begin{subfigure}[b]{.5\linewidth}
      \centering
      \includegraphics[width=.90\linewidth]{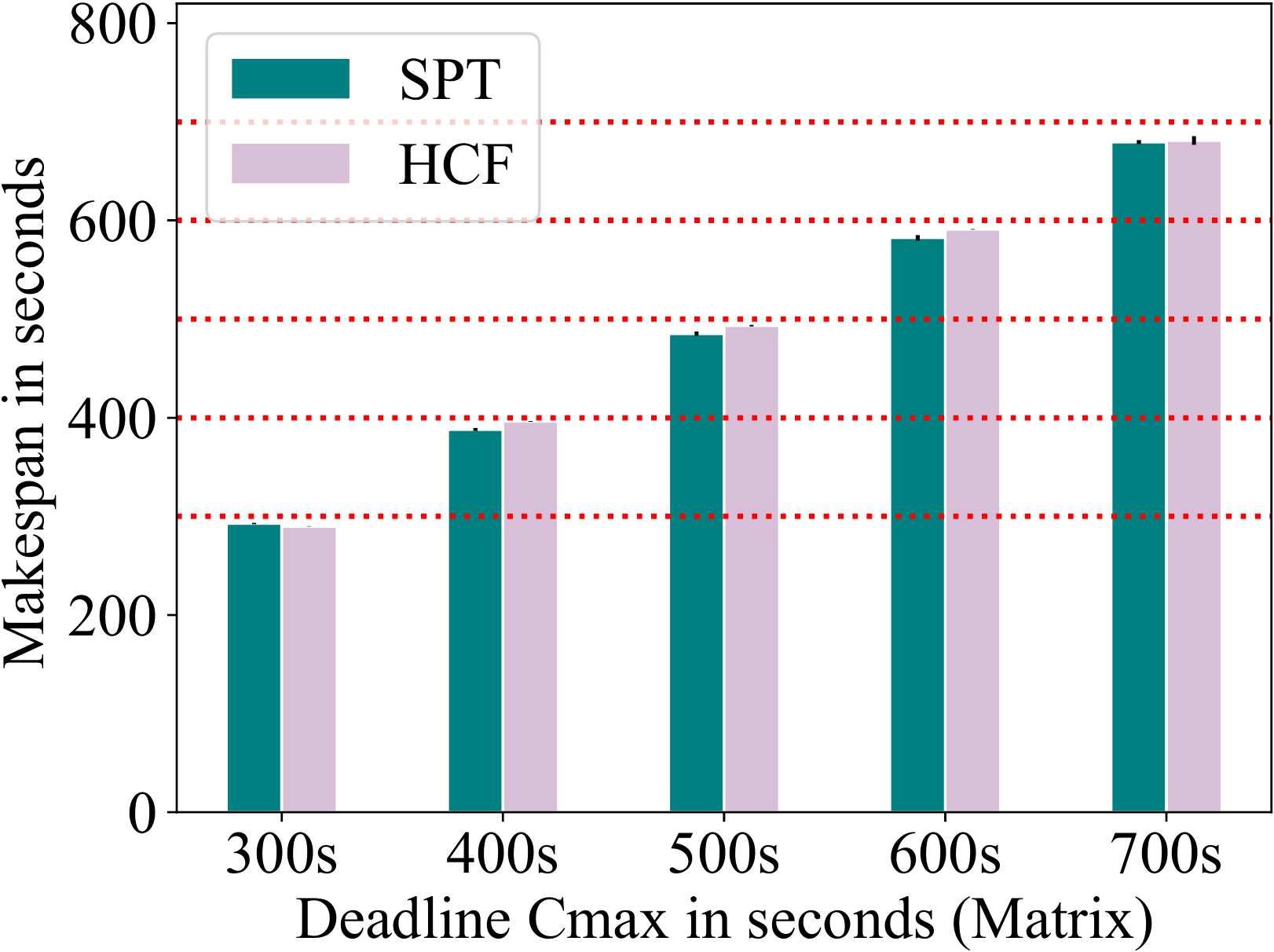}
      \caption{Matrix Processing}      
      \label{fig.all_jobs_matrix_makespan}
     \end{subfigure}%
\begin{subfigure}[b]{.5\linewidth}
      \centering
      \includegraphics[width=.90\linewidth]{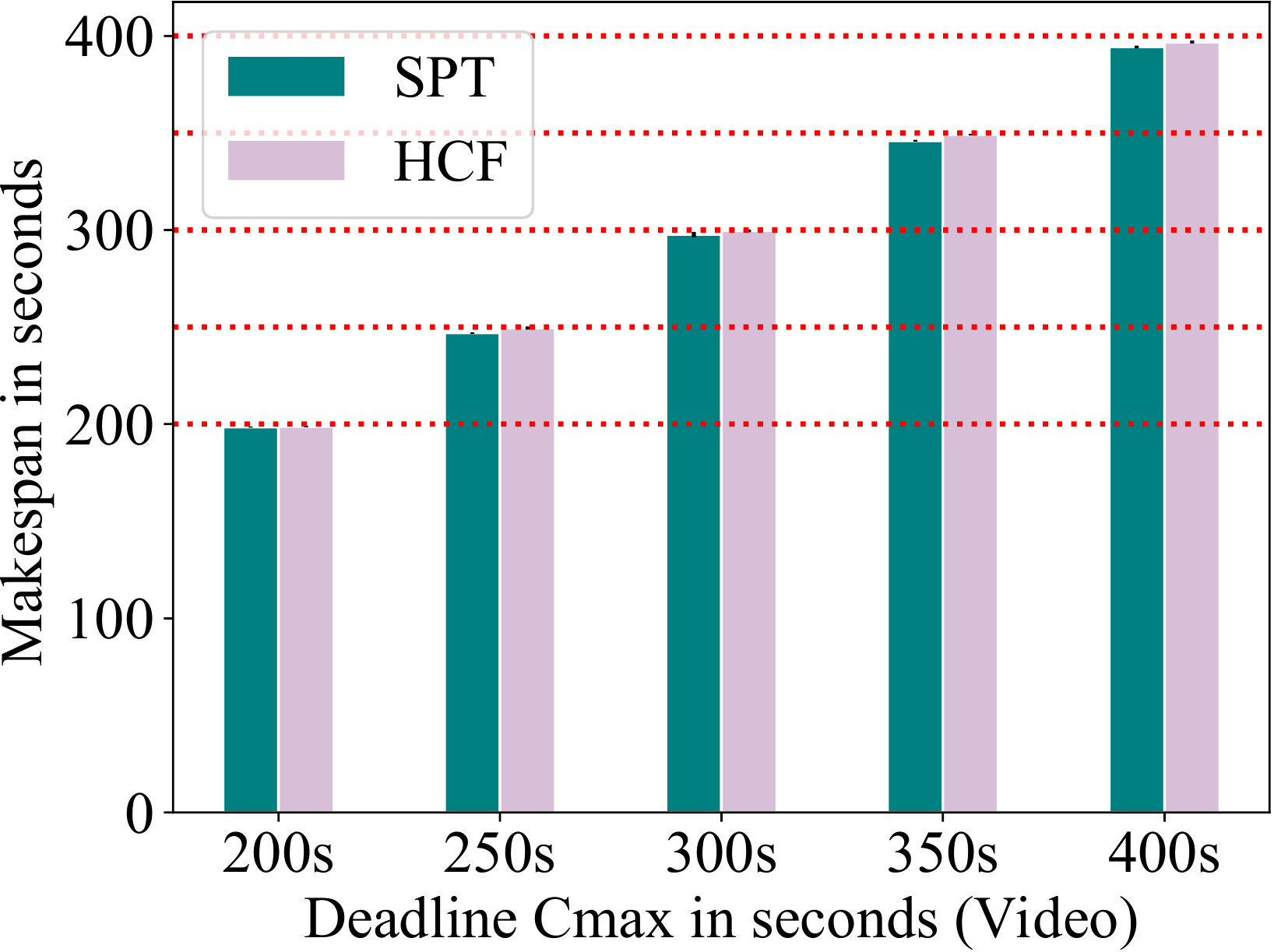}
      \caption{Video Processing}
      \label{fig.all_jobs_video_makespan}
     \end{subfigure}
\caption{Actual execution makespan for the SPT and HCF heuristics with varying $C_{max}$ for the Matrix and Video Processing applications.}
\label{fig.all_jobs_cmax_variation_makespan}
\squeezeuppicture
\end{figure}

In Fig.~\ref{fig.all_jobs_matrix_makespan} and \ref{fig.all_jobs_video_makespan}, we show the actual obtained makespan from our hybrid scheduling framework for the Matrix and Video processing applications with varying $C_{max}$.  
In both the applications, we find that the observed makespan is very close to the user desired value of $C_{max}$, with $<$~3.5\% and $<$~1.5\% absolute error, respectively for the Matrix and Video Processing applications. This shows the validity of the performance of our scheduler and the heuristics. Prediction errors in the performance models contributes a lot towards this error during the scheduling, as having inaccurate prediction models for $P_{k,j}^{private}$ would prevent the scheduler from utilizing the private cloud at the highest possible efficiency. 


\sep{This paragraph and the next both start with 'we futher note'}
In our experiments, the all-private cloud execution has the makespan of 740~s for executing all 150 jobs in Matrix Processing application and 407~s for all 200 jobs in the Video Processing application. Therefore, given a reasonable deadline of $C_{max}=400~s$ for Matrix Processing application and $C_{max}=250~s$ for Video Processing application, our framework via SPT ordering can achieve a speedup of 1.92$\times$ and 1.65$\times$ over an approach that uses only private cloud execution, at a cost which is, respectively, 40.5\% and 39.5\% of an approach that uses only the public cloud.

We further note that for applications like the Image Processing, where the compute latencies are the order of 100s of milliseconds, the effects of error in scheduling will be much larger as communication and coordination latencies between different parts of intra and inter public-private clouds can introduce large variances. However, the absolute error in makespan with SPT ordering is $\approx$ 5\% and that with HCF ordering is $\approx$ 23\%, which is comparable to the error we get in our prediction models. This suggests that the scheduling framework can perform with higher accuracy for moderate to heavy workloads.


\section{Related Work} \label{related.sec}
Various approaches for computational offloading and scheduling in the context of datacenters, microservices, and serverless architectures have been proposed in recent years. 
In context of mobile cloud computing, computational offloading or cyber-foraging using virtual machines has been studied extensively
\cite{survey_opportunistic_offloading}, \cite{mach2017mobile}. Here, tasks can be offloaded from resource constrained mobile devices as needed to meet performance objectives such as minimizing execution cost, energy usage or latency. The offloading problems are generally solved using methods such as integer liner programs, greedy heuristics, or dynamic programming. 
A similar work to our is DEFT~\cite{deft}, which uses regression-based performance models to dynamically determine where to offload computation to optimize latency or energy. However, DEFT offloads entire single-stage applications, whereas our framework makes finer grained, per stage offloading decisions to optimize cost.
In context of serverless computing, more recently, several systems have been proposed for scheduling single stage serverless functions in a single cloud platform. 
Spock~\cite{spock} minimizes service-level objective violations by distributing the execution of machine learning inference jobs over serverless functions and VMs in the public cloud. The goal of the FnSched scheduler~\cite{suresh2019fnsched} is to maximize resource utilization from a cloud 	platform provider's perspective while meeting latency guarantees. This is done via regulating the allocated resources to the function containers based on their resource consumption patterns. 
NOAH~\cite{stein2018serverless} is a framework that uses a game theoretic approach for scheduling and resource allocation of single stage functions in a private cloud to minimize response time.

Multi-stage serverless applications have been studied in~\cite{baldini2017serverless}, 
where the authors present an implementation of function chaining in Apache OpenWhisk. The authors in \cite{ao2018sprocket} propose a multistage serverless video processing framework. 
The authors in~\cite{edgetaskplacement} studied task placement of multi stage applications in edge-cloud systems to minimize completion time, however they do not consider actual cost, and schedule on a per input basis, whereas we consider a batch input. In~\cite{elgamal2018costless}, the authors propose Costless, a framework that optimizes the cost of serverless application execution by splitting a function chain and executing part on an edge platform and part on the cloud. Costless optimizes for a single application execution at a time and does not consider sequencing and scheduling tasks concurrently in function replicas.
The authors in~\cite{kannan2019grandslam} propose GrandSLAm, a framework for scheduling machine learning workloads to maximize data center resource utilization by dynamically adjusting  batch sizes and and reordering the executing of requests.
In contrast to these works, we focus on task scheduling in a hybrid cloud setting. In addition, our approach optimizes for performance and cost for the platform clients, rather than for the service provider.

%

\section{Conclusion} \label{conclusion.sec}
We have presented a framework for scheduling serverless applications over a hybrid public-private cloud in a manner that minimizes the cost of public cloud use, while meeting a user-specified makespan constraint. 
We proved this problem to be $\mathcal{NP}$-hard and proposed a greedy algorithm with two heuristics. We then  presented the details of our framework, which relies on accurate predictive models for function compute time, intermediate data sizes, and network transfer latencies. 
Finally, we presented an evaluation of a prototype implementation of our framework that uses AWS Lambda for the public cloud and OpenFaaS, running in an on-premise server, for the private cloud using canonical examples of serverless applications. 
Our results showed that our framework can achieve a speedup of $1.92$ times in the Matrix Processing application and $1.65$ times in the Video Processing application over an approach that uses only the private cloud, at a cost that is, respectively,  $40.5$ percent and $39.5$ percent of an approach that uses only the public cloud. Our framework essentially handles each application stage independently, hence, we believe it is possible to extend this decoupled approach to complicated DAGs. In future work, we plan on extending our approach to introduce a dynamic tolerance to the deadline violation and to minimize a dual cost / makespan objective. 


\section*{Acknowledgment}
This work is supported by the National Science Foundation under grants CNS 1553340 and CNS 1816307, Air Force Office of Scientific Research (AFOSR) under grant FA9550-19-1-0054, and an AWS Cloud Credits for Research grant.

\bibliographystyle{IEEEtran}
\bibliography{biblio}

\appendix[Problem Formalization] \label{problem.sec}

In this section, we present a formal definition of the hybrid cloud scheduling problem. We also formulate the problem into a MILP which could be solved using a standard solver.

\begin{table}[htpb]
\caption{Known constants and problem inputs.} \label{table.knownvar}
\normalsize
\centering
\resizebox{.9\linewidth}{!}{
\begin{tabular}{|l|l|}
\hline
\textbf{Notation} & \textbf{Description of known quantities} \\ \hline
$P_{k,j}^{public}$ & Latency for execution of stage $k$ of job \\ & $j$ in public cloud including transmission \\
& time and startup latency. \\ \hline
$P_{k,j}^{private}$ & Latency for execution of stage $k$ for \\ & job $j$ in the private cloud. \\ \hline
$H_{k,j}$ & Cost of executing the $k^{th}$ stage of job\\ & in the public cloud. \\ \hline
$D_{k,j}$ & Latency of downloading results of $k^{th}$ \\& stage of job from the public cloud. \\ \hline
$U_{k,j}$ & Latency of uploading results of $k^{th}$ \\ & stage of job to the public cloud. \\ \hline
$I_k$ & \# of private cloud replicas for stage $k$. \\ \hline
$\mathcal{G} = (V_j, E_j)$ & DAG for job $j$. \\ \hline
$\Omega_j$ & Set of functions for job $j$ that must be \\&  executed in the private cloud. \\ \hline
$C_{max}$ & Deadline to complete all jobs. \\ \hline
$\delta_k$ & The out degree of node for stage $k$\\
\hline
\end{tabular}
}
\end{table}

\begin{table}[htpb]
\normalsize
\centering
\caption{Decision variables.} \label{table.decisionvar}
\resizebox{.9\linewidth}{!}{
\begin{tabular}{|l|l|}
\hline
\textbf{Notation} & \textbf{Description of decision variables} \\ \hline
$\skj{k}{j}$ & Start time of stage $k$ of job $j$, $j = 1 \ldots J$, \\
& $k \in V_j$.  \\ \hline
$\xkji{k}{j}{i}$ & 1 if job $j$ is processed at replica $i$ for \\
& stage $k$, else 0. \\ \hline
$e_{k,j}$ & 0 if stage $k$ of job $j$ executes in the public \\
& cloud, 1 if stage $k$ of job $j$ executes in the \\ & private cloud. \\ \hline
$\ykjr{k}{j}{r}$ & 1 if job $j$ precedes job $r$ in stage $k$. \\ \hline
$u_{k,j}$ & 1 if result of stage $k$ needs to be uploaded \\
& to cloud for job $j$. \\ \hline
$d_{k,j}$ & 1 if result of stage $k$ needs to be downloaded\\
& from cloud for job $j$. \\ \hline
$X_k$ & $\delta_p \times e_{p,j} - \sum\limits_{(p,q) \in E_j} e_{q,j}$, $j = 1 \ldots J$, $k \in V_j$ \\ \hline
\end{tabular}
}
\end{table}

\subsection{Problem Specification}

We formalize the scheduling problem as a variant of a flow shop scheduling problem~\cite{emmons2012flow}.
The constants and inputs to the problem are shown in Table~\ref{table.knownvar}. Of note, we assume that the latencies $P_{k,j}^{public}$ and $P_{k,j}^{private}$  for executing each function in the public and private cloud, respectively, are known. The latencies are input-dependent and thus vary from job to job.  In practice, we obtain these latencies using performance models. 
The  cost for executing a function in the public cloud  $H_{k,j}$ is computed as function of the execution latency according to Equation  \eqref{eqn.cost}. 
We also note that the application owner can dictate which stages must only be run on the private cloud,  for privacy or security reasons, for example. These constraints are specified on a per job basis. Each job $j$ in a single application follows the DAG $\mathcal{G}$, where the stages of the DAG are represented by $V_j$ and are same for all jobs in that application. 

Table~\ref{table.decisionvar} shows the decision variables for the scheduling problem. 
The solution specifies, for each stage of each job, whether it is executed in the private cloud ($e_{k,j}=1$) or the public cloud ($e_{k,j}=0$).
For each stage of a job that is executed in the private cloud, the solution dictates in which replica the stage will execute.
The solution also specifies the start time for the execution of each stage. Thus, the solution fully describes a \emph{schedule} for all stages of all jobs in the workload. 

The objective of the scheduling problem is to minimize the total cost of using the public cloud, subject to the constraint that all 
jobs finish within $C_{max}$ time.  To formalize this objective in the canonical form, we instead seek to maximize the cost saved by the stages that are executed in the private cloud. This total cost savings is
\[
z = \sum \limits_{k \in V_j } \sum \limits_{j=1}^J e_{k,j} \times H_{k,j},
\]

where $e_{k,j} = 1$ if stage $k$ of job $j$ is executed in the private cloud (and 0 otherwise), and $H_{k,j}$ is the cost that would have been incurred if this stage was executed in the public cloud, i.e., the cost saved by executed the stage in the private cloud.

\subsection{Mixed-Integer Linear Program Formulation}
Here, we formulate a MILP for the hybrid cloud scheduling problem:
\begin{align}
& \text{maximize} \quad  z = \sum_{k \in V_j} \sum \limits_{j=1}^J e_{k,j} \times H_{k,j} \label{objective.eq} \\
& \textrm{subject to} \nonumber \\
& \skj{m}{j} + \pkj{m}{j}^{private}\times e_{m,j} + \pkj{m}{j}^{public}\times (1-e_{m,j}) \leq C_{max}, \nonumber \\
& \quad \quad \quad  \quad j = 1 \dots J; m \in {V_j} \label{makespan.con}  \\
& \skj{q}{j} \geq \skj{p}{j} + \pkj{p}{j}^{private}\times e_{k,j} + \pkj{p}{j}^{public} \times (1-e_{k,j}) \nonumber \\
& \quad \quad + u_{p,j} \times U_{p,j} + d_{p,j} \times D_{p,j}, \nonumber \\
& \quad \quad \quad  \quad j=1 \ldots J; (p,q) \in E_j \label{dag.con}  \\
& \sum \limits_{i \in I_k} \xkji{k}{j}{i} = e_{k,j}, \quad j = 1 \ldots J; k \in V_j  \label{pod.con} \\ 
& \skj{k}{j} - (\skj{k}{r} + \pkj{k}{r}^{private}) + Q(2 + \ykjr{k}{j}{r} - \xkji{k}{j}{i} - \xkji{k}{r}{i}) \geq 0 , \nonumber \\
& \quad \quad \quad \quad j,r =1 \dots J : j < r; k \in V_j; i \in I_k \label{schedule1.con} \\
& \skj{k}{r} - (\skj{k}{j} + \pkj{k}{j}^{private}) + Q(3 - \ykjr{k}{j}{r} - \xkji{k}{j}{i} - \xkji{k}{r}{i}) \geq 0 , \nonumber \\
& \quad \quad \quad \quad  j,r =1 \dots J : j < r; k \in V_j; i \in I_k  \label{schedule2.con} \\
& X_k \geq 0.001 - M(1 - U_{k,j}) \;, j=1\cdots J, k \in V_j \label{startupdown.con}\\ 
& X_k \leq MU_{k,j} \;, j=1\cdots J, k \in V_j \\
& X_k \leq - 0.001 + M(1 - D_{k,j})  \;, j=1\cdots J, k \in V_j\\
& X_k \geq -MD_{k,j} \;, j=1\cdots J, k \in V_j\label{stopupdown.con}\\
& e_{k,j} =1, \quad j=1 \ldots J; k \in \Omega_j \label{private.con} \\
& \skj{k}{j} \geq 0, \quad \forall j = 1,2,\cdots J; k \in V_j  \label{start.con} \\
& e_{k,j} \in \{0,1\}, \quad j = 1,2,\cdots J; k \in V_j   \label{privatebin.con} \\
& \ykjr{k}{j}{r} \in \{0,1\}, \quad j,r =1 \ldots J; k \in V_j  \label{ybin.con} \\
& \xkji{k}{j}{r} \in \{0,1\} , \quad  j,r =1 \ldots J; k \in V_j.  \label{xbin.con}
\end{align}

As stated above, the objective (\ref{objective.eq}) is to maximize the cost savings of stages executed in the private cloud.
The constraint (\ref{makespan.con}) requires that all jobs complete within the makespan $C_{max}$.
Constraint (\ref{dag.con}) ensures that the start time of stage $q$ of job $j$ is after the finishing time of stage $p$ of job $j$ in if there exists a directed edge $(p,q)$ in $\mathcal{G}_j$. Constraint~(\ref{pod.con}) ensures that if stage $k$ of job $j$ executes in the private cloud, it is assigned to exactly one replica.
Constraints (\ref{schedule1.con}) and (\ref{schedule2.con}) determine the sequence of job stages on each replica; for any two job stages assigned to the same replica, one job must finish before the subsequent job can start. Here, $Q$ is a very large integer, and M is another large integer; their use serves the purpose of conditionals. Either constraint (\ref{schedule1.con}) or constraint (\ref{schedule2.con}) applies, but not both, depending on the evaluation of the second term in each inequality. 
Constraint (\ref{private.con}) ensures that job stages that are required to run in the private cloud are scheduled to do so. Constraints (\ref{startupdown.con}-\ref{stopupdown.con}) handles the upload and download latencies.
Constraint (\ref{start.con}) requires that the start time of each stage of each job is non-negative, and 
(\ref{privatebin.con}) - (\ref{xbin.con}) constrain the integer decision variables to be in the set $\{0,1\}$.
%

\textit{Special Case:} When there is a single stage for all jobs, i.e., $|V_j| = 1, \; \forall j$, then the scheduling problem reduces to a multiple knapsack problem, where the size of each knapsack is $C_{max}$. There are $I_1$ knapsacks in the first and only stage. The scheduling problem then reduces to the problem of fitting as many of the most expensive jobs  in $I_1$ replicas within the deadline $C_{max}$. The jobs outside the knapsack are the jobs that are  offloaded to the public cloud.  

\subsection{Proof of $\mathcal{NP}$-Hardness} 
We now prove that the hybrid cloud scheduling problem is $\mathcal{NP}$-hard. To do this, we show that the decision version of the problem is $\mathcal{NP}$-complete. We denote the latency for execution of stage $k$ of job $j$ in public cloud including transmission time and startup latency by $P_{k,j}^{public}$, and  
corresponding execution latency in the private cloud by $P_{k,j}^{private}$.

\begin{problem}[The Hybrid-Cloud Scheduling Decision Problem] \label{hybriddecision.prob}
 Given a set of input jobs, a hybrid platform with a fixed allocation of private cloud replicas per stage, a target maximum makespan $C_{max}$, and a target budget $B$, is there a schedule with makespan less than or equal to $C_{max}$ that has public cloud cost less than or equal to $B$?
\end{problem}
\begin{theorem}
The Hybrid Cloud Scheduling Decision Problem is $\mathcal{NP}$-complete.
\end{theorem}
\begin{IEEEproof}
First, we note that the problem is in $\mathcal{NP}$. Given a schedule, we can verify in polynomial time that its makespan and public cloud cost are less than or equal to $C_{max}$ and $B$, respectively. 

We now show that we can reduce a known $\mathcal{NP}$-hard problem, the flow shop scheduling problem with three machines, $F3||C_{max}$~\cite{emmons2012flow}, to our decision problem. 
In $F3||C_{max}$, there is set of $n$ input jobs. Each job must execute on each of three machines $m_1$, $m_2$, and $m_3$, in that order, and each job has a known processing time per machine.
The $F3||C_{max}$ decision problem is to determine whether there is a schedule that assigns the jobs to machines so that the  makespan (defined as the total time required to execute all jobs on all machines) is less than or equal to $C_{max}$.

Given an instance of $F3||C_{max}$, we transform it to an instance of Problem~\ref{hybriddecision.prob} as follows. We consider an application with three stages, where each stage has a single replica, i.e., $I_k =1$, for $k=1,2,3$.  For each input job to $F3||C_{max}$, we create an input job to our problem, with the execution time for job $j$ in the private cloud $P_{k,j}^{private}$ equal to the execution time of job $j$ on machine $k$ in $F3||C_{max}$, for $k=1,2,3$. The execution time in the public cloud $P_{k,j}^{public}$ is set to an arbitrary  positive number. Further, the cost for executing in the public cloud $H_{k,j}$ is set to 1 for all jobs $j$ and $k=1,2,3$. The DAG for each job $j$ is defined naturally to mandate that first stage~1 executes, followed by stage~2, and then by stage~3. For all jobs $j$, the set of all functions that must execute in the private cloud is $\Omega_j = \{1,2,3\}$, i.e., all jobs must execute all stages in the private cloud. Further, we use $C_{max}$ for $F3||C_{max}$ as $C_{max}$ for our decision problem,  and we set $B = 0$. It is straightforward to observe from the construction, that the solution to  Problem~\ref{hybriddecision.prob} is $YES$ if and only if the solution to $F3||C_{max}$ is $YES$.
\end{IEEEproof}

\end{document}